\documentclass[]{mn2e}
\usepackage[dvips]{graphicx}
\usepackage{times}
\usepackage{times,amssymb}
\newcommand{\be}{\begin{equation}}
\newcommand{\ee}{\end{equation}}
\newcommand{\bea}{\begin{eqnarray}}
\newcommand{\eea}{\end{eqnarray}}
\begin{document}

\title[Dynamics of two planets in co-orbital motion]
      {Dynamics of two planets in co-orbital motion}

\author[C.A. Giuppone, C. Beaug\'e, T.A. Michtchenko and S. Ferraz-Mello]
{C.A. Giuppone$^1$, C. Beaug\'e$^1$, T.A. Michtchenko$^2$ and S. Ferraz-Mello$^2$\\
$^{1}$Observatorio Astron\'omico, Universidad Nacional de C\'ordoba, Laprida 854, 
(X5000BGR) C\'ordoba, Argentina\\
$^{2}$Instituto de Astronomia, Geof\'{\i}sica e Ci\^encias Atmosf\'ericas, USP, Rua 
do Mat\~ao 1226, 05508-900 S\~ao Paulo, Brazil}

\date{}

\pagerange{\pageref{firstpage}--\pageref{lastpage}} \pubyear{2009}

\maketitle

\label{firstpage}

\begin{abstract}
We study the stability regions and families of periodic orbits of two planets locked 
in a co-orbital configuration. We consider different ratios of planetary masses and 
orbital eccentricities, also we assume that both planets share the same orbital 
plane. Initially we perform numerical simulations over a grid of osculating initial 
conditions to map the regions of stable/chaotic motion and identify equilibrium 
solutions. These results are later analyzed in more detail using a semi-analytical model.

Apart from the well known quasi-satellite (QS) orbits and the classical equilibrium Lagrangian points $L_4$ and $L_5$, we also find a new regime of asymmetric periodic solutions. For low eccentricities these are located at
$(\Delta \lambda,\Delta \varpi) = (\pm 60^\circ,\mp 120^\circ)$, where $\Delta \lambda$ is the difference in mean longitudes and $\Delta \varpi$ is the difference in longitudes of pericenter. The position of these {\it Anti-Lagrangian} solutions changes with the mass ratio and the orbital eccentricities, and are found for eccentricities as high as $\sim 0.7$. 

Finally, we also applied a slow mass variation to one of the planets, and analyzed its effect on an initially asymmetric periodic orbit. We found that the resonant solution is preserved as long as the mass variation is adiabatic, with practically no change in the equilibrium values of the angles. 
\end{abstract}


\section{Introduction}
In the restricted three-body problem there are different domains of stable motion associated to co-orbital motion. Each can be classified according to the center of libration of the critical argument, $\sigma = \lambda -\lambda$', where $\lambda$ denotes the mean longitude of the minor body and $\lambda$' the same variable for the disturbing planet. These types of motion are known as: {\it (i)} Tadpole orbits, corresponding to a libration of $\sigma$ around $L_4$ or $L_5$; {\it (ii)} Horseshoe orbits, where motion occurs around $\sigma = 180^{\circ}$ and encompasses both equilateral Lagrangian Points, and {\it (iii)} quasi-satellite (QS) orbits, where $\sigma$ oscillates around zero. 

The term ``quasi-satellites'' was originally introduced by Mikkola and Innanen (1997) and can be viewed as an extension of retrograde periodic orbits in the circular restricted three-body problem (e.g. Jackson 1913, H\'enon 1969). Although not present for circular orbits, they exist for moderate to high eccentricities of the particle. In a reference frame rotating with the planet, QS orbits circle the planet like a retrograde satellite, although at distances so large that the particle is not gravitationally bounded to the planetary mass (Mikkola et al 2006).

The first object confirmed in a QS configuration was the asteroid $2002 VE68$ (Mikkola et al., 2004) with Venus as the host planet. The Earth has one temporary co-orbital object, ($3753 \, Cruithne$, Namouni et al. 1999), and one alternating horseshoe-QS object ($2002 AA29$, Connors et al. 2002). The co-orbital asteroidal population in the inner Solar System was studied in Brasser et al. (2004) by numerical integrations. All QS orbits appear to be temporary, escaping in timescales of the order of $10^2-10^4$ years

Wiegert et al. (2000) numerically investigated the stability of QS orbits around the giant planets of the Solar System. Although no stable solutions were found for Jupiter and Saturn, some initial conditions around Uranus and Neptune lead to QS orbits that survive for timescales of the order of $10^9$ yr. It thus appears that a primordial population of such objects may still exist in the Solar System. Kortenkamp (2005) used N-body simulations to model the combined effects of solar nebula gas drag and gravitational scattering of planetesimals by a protoplanet. He showed that a significant fraction of scattered planetesimals could become trapped into QS trajectories. It then seems plausible that this trapped-to-captured transition may be important not only for the origin of captured satellites but also for continued growth of protoplanets.

At variance with these results, in the case of the general (non-restricted) three-body problem, although equilateral solutions and horseshoe orbits are well known, quasi-satellite configurations have only been studied very recently. Hadjidemetriou et al. (2009) performed a detailed study of periodic orbits in the 1/1 MMR for fictitious planetary systems with different mass ratios. They found that stable QS solutions occur for $\sigma = \Delta \lambda = \lambda_2 - \lambda_1 = 0$ and $\Delta\varpi = \varpi_2 - \varpi_1 = 180^{\circ}$, where the subscripts identify each planet. Unstable trajectories were found at $\sigma = 180^{\circ}, \Delta\varpi = 0$. Although at present there are no confirmed cases of exoplanets in quasi-satellite configurations, Go\'zdziewski \& Konacki (2006) found that the radial velocity curves of the HD82943 and HD128311 planets could correspond to co-orbital motion in highly inclined orbits. Numerical simulations of both systems show QS trajectories, instead of Trojan orbits as initially believed. 

In the present work we aim to revisit the 1/1 mean-motion resonance (MMR) in the planar planetary three-body problem, trying to identify possible domains of stable solutions and their location in the phase space. Section \ref{regular} presents several dynamical maps constructed from numerical simulations for different initial conditions. These maps allow us to identify stable fixed points and periodic orbits, as well as the domains of regular motions. In Section \ref{model} we develop a semi-analytical model for co-orbital planets, which is then applied in Section \ref{op} to calculate the families of stable periodic orbits. In the same section we also present a brief study of the effects of an adiabatically slow mass variation in one of the planetary bodies. Finally, conclusions close the paper in Section \ref{conclusions}.

\begin{figure}
\centerline{\includegraphics*[width=18pc]{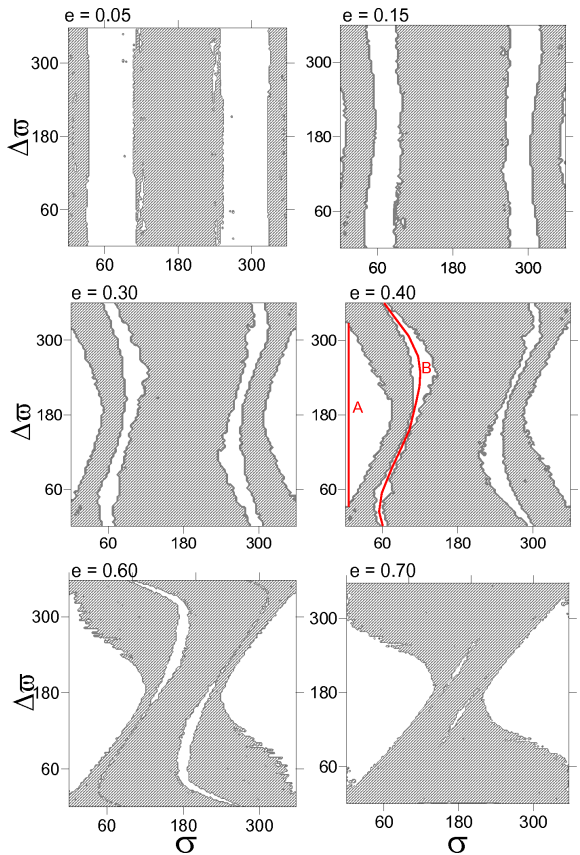}}
\caption{Results of numerical integrations of initial conditions in a grid in the $(\sigma, \Delta \varpi)$ plane. Planetary masses were taken equal to $m_1=m_2= m_{\rm Jup}$, and initial semimajor axes equal to $a_1 = a_2 = 1$ AU. Regions of regular motion are shown in white, while the dashed regions correspond to chaotic and unstable trajectories.}
\label{fig1}
\end{figure}

\begin{figure}
\centerline{\includegraphics*[width=18pc]{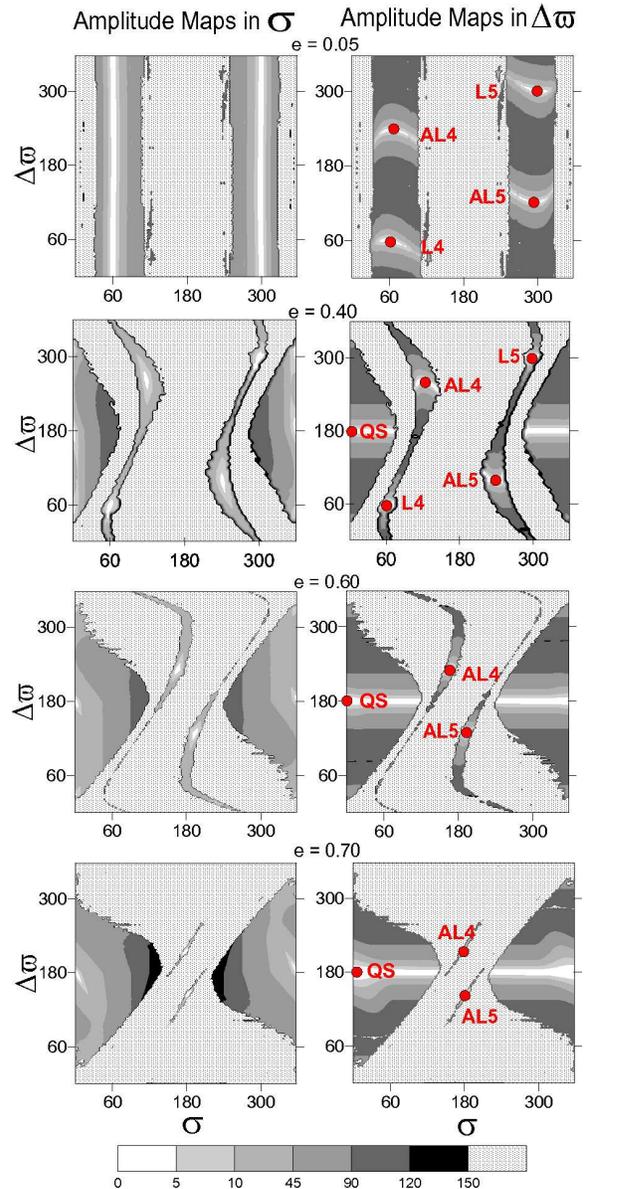}}
\caption{Semi-amplitude maps. The Left (Right) column shows the amplitude variation for $\sigma$ ($\Delta\varpi$) in gray scale. Light domains correspond to near zero amplitude families, darker regions indicate oscillation amplitudes up to $90^{\circ}$, and dashed regions correspond to unstable orbits. Initial values of eccentricities are indicated in each panel. Color scale is indicated at bottom and ACR solutions are marked on the right panels.}
\label{fig2}
\end{figure}

\section{Dynamical Maps with Equal Mass Planets}\label{regular}

Consider two planets with masses $m_1$ and $m_2$ in coplanar orbits around a star with mass $m_0=M_{\odot}$. We will begin considering the case $m_2=m_1$, also other mass ratios will be discussed in later sections. Let $a_i$ denote the semimajor axes, $e_i$ the eccentricities, $\lambda_i$ the mean longitudes and $\varpi_i$ the longitudes of pericenter. All orbital elements considered in this paper are assumed astrocentric and osculating. Throughout this work, $m_1$ will be our ``reference'' planet: its mass will be fixed at one Jovian mass ($m_1 = M_{\rm Jup}$) and the system scaled to initial condition $a_1 = 1$ AU. The angular variables for co-orbital motion will then be defined as $\sigma = \lambda_2-\lambda_1$ and $\Delta\varpi = \varpi_2-\varpi_1$. 

As pointed out by Hadjidemetriou et al. (2009), for equal mass planets the periodic orbits are such that are located at $a_1=a_2$ and $e_1=e_2$. Accordingly, we fixed the semimajor axes and eccentricities, and constructed a $100 \times 100$ grid of initial conditions varying both $\sigma$ and $\Delta\varpi$ between zero and $360^{\circ}$. Each point in the grid was then numerically integrated over $3000$ orbital periods using a Bulirsch-Stoer based N-body code, and we calculated the averaged MEGNO chaos indicator $ \langle Y \rangle$ (Cincotta \& Sim\'{o}, 2000) to identify regions of regular or chaotic motion. Results are shown in Figure \ref{fig1} for six values of the initial eccentricities $e_i$; dashed regions correspond to unstable orbits while white was used to identify stable solutions. An analysis of these plots show the following characteristics:

\begin{itemize}

\item For low initial eccentricities ($e_i=0.05$) the maps show two disconnected strips of regular motion, corresponding to motion around $\sigma = \pm 60^{\circ}$ and any value of $\Delta\varpi$.

\item For moderate low to intermediate initial eccentricities ($e_i=0.15$ and $e_i=0.30$) the vertical strips of regular motion become thinner and slightly distorted. A new stable domain is now present, associated to QS orbits, and located around $\sigma=0$. 

\item For high initial eccentricities ($e_i \ge 0.40$) the domain of QS orbits increases and covers a significant portion of the plane of initial conditions. Conversely, the distorted vertical strips shrink and each seems to break into two islands of stable motion. The smaller islands encompass equilateral solutions, although they almost disappear for $e_i=0.70$. The larger islands correspond to a different type of asymmetric solution, and their locations tend towards the center of the plots as the eccentricities increase.

\item Due to symmetry present in the dynamical system, the results are invariant to transformations of the type $(\sigma,\Delta \varpi) \rightarrow (-\sigma,-\Delta \varpi)$. In fact, since $m_1=m_2$, both equilateral solutions are actually the same solution, since we can pass from one to the other just by redefining the reference planet. However, since later sections will discuss the case $m_2 \ne m_1$, we prefer to treat both equilateral solutions separately.

\end{itemize}

Although MEGNO is a very efficient tool to identify chaotic motion, it is not suited to distinguish between different types of regular orbits (e.g. fixed points, periodic orbits, etc.). Sometimes this task is performed with a Fourier transform of the numerical data (e.g. Michtchenko et al. 2008ab); however, here we have chosen a different route. Starting from the output of each numerical simulation, we calculated the amplitudes of oscillation in each angular variable. Initial conditions with zero amplitude in $\sigma$ correspond to {\it $\sigma$-family} periodic orbits of the co-orbital system, while solutions with zero amplitude in $\Delta \varpi$ will correspond to periodic orbits of the so-called {\it $\Delta \varpi$-family} (see Michtchenko et al. 2008ab). Finally, stationary solutions of the averaged problem, identified as intersections of both families, may be thought as analogous to the apsidal corotation resonances (ACR) found in other mean-motion resonances (e.g. Beaug\'e et al. 2003). The equilateral Lagrangian solutions will appear as ACR in these plots. 

The gray scale graphs in Figure \ref{fig2} show values of the amplitudes in $\sigma$ (left) and $\Delta \varpi$ (right) for four of the plots shown in figure \ref{fig1}. White regions represent initial conditions with semi-amplitudes smaller than $2^{\circ}$, as thus indicate the families of periodic orbits in each angle. Darker regions correspond to increasing amplitudes (up to $45^{\circ}$) and denote initial conditions with quasi-periodic motion. The dashed areas are unstable solutions. 
Finally, it is worthwhile mentioning that symmetric configurations may either correspond to an alignment ($\Delta\varpi= 0^{\circ}$) or an antialignment of the apses ($\Delta\varpi= \pm 180^{\circ}$) while asymmetric configurations have stationary values of $\Delta\varpi$ different from the above.

For low eccentricities ($e_i=0.05$) we observe four asymmetric ACR solutions. Two are the well known Lagrangian equilateral solutions located at $(\sigma,\Delta \varpi) = (\pm 60^{\circ},\pm 60^{\circ})$. By analogy with the restricted problem, we will denote them $L_4$ and $L_5$. As far as we know, the remaining two ACR have not been previously reported, and are located at approximately $(\sigma,\Delta \varpi) = (\pm 60^{\circ},\mp 120^{\circ})$. We have called them {\it Anti-Lagrangian solutions} and they are connected to the classical equilateral Lagrangian solutions by the $\sigma$-family of periodic orbits. By analogy, we have denoted the new solutions as:
\bea
\label{eq1}
AL_4 &:& \sigma \in [0,180^\circ] \hspace*{0.9cm} \Delta \varpi \in [180^\circ,360^\circ] \\
AL_5 &:& \sigma \in [180^\circ,360^\circ] \hspace*{0.43cm} \Delta \varpi \in [0^\circ,180^\circ]   \nonumber .
\eea
As with all previous stationary solutions, these asymmetric points are found at $a_1=a_2$. 

\begin{table}
\begin{center}
\begin{tabular}{|c|c|c|c|c|}
\hline 
&  $\sigma$ (deg) & $\Delta\varpi$ (deg)    \\
\hline
 QS       &    0   &   180    \\
 $L_4$    &   60   &    60    \\
 $L_5$    &  300   &   300    \\
 $AL_4$   &   60   &   240    \\
 $AL_5$   &  300   &   120    \\
\hline 
\end{tabular} 
\end{center}
\caption{Approximate location for the stable ACR solutions in the $(\sigma, \Delta \varpi)$ plane, calculated from the dynamical maps with $e_1=e_2=0.15$. For equal mass planets, all stationary solutions occur for $a_1=a_2$.}
\label{tab_pos}
\end{table}

As the eccentricities grow (e.g. $e_i=0.40$) the QS region at $(\sigma,\Delta \varpi)=(0,180^\circ)$ causes a distortion and compression of the stable asymmetric domain. The  Anti-Lagrangian zone seems less affected and surrounded by a larger island of stable motion. This effect is even more pronounced for $e_i=0.60$ and $e_i=0.70$ where the stable domain around $L_4$ and $L_5$ almost disappear. The region around $AL_4$ and $AL_5$ are still visible, although they also decrease in size and their location approaches the unstable symmetric periodic orbit located at $\sigma = \Delta \varpi = 180^\circ$. 

The decrease in the size of the stable regions around the asymmetric ACR solutions is accompanied by a significant increase in the stable domains around QS orbits, which, for high eccentricities, seem to cover a large proportion of the plane. Inside this region we also note two families of periodic orbits; the $\Delta \varpi$-family which is restricted to a small region around $\Delta \varpi = 180^\circ$, and a smaller $\sigma$-family close to zero value of the resonant angle. 

\begin{figure}
\centerline{\includegraphics*[width=20pc]{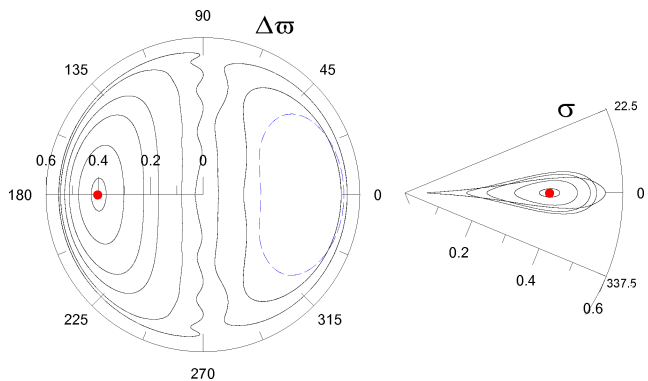}}
\caption{Variation of the eccentricity of each planet with $\Delta\varpi$ (left frame) and with $\sigma$ (right frame). Initial conditions were chosen inside the QS region following line A in Fig. \ref{fig1} for $e=0.40$. The radial distance is the value of the osculating eccentricity. The position of the ACR solution is shown in red, and is located at $\Delta\varpi=180^\circ$. Note, however, the existence of large-amplitude solutions around $\Delta\varpi=0$, even though no stable ACR solution is found in this region.}
\label{fig3}
\end{figure} 

\begin{figure}
\centerline{\includegraphics*[width=20pc]{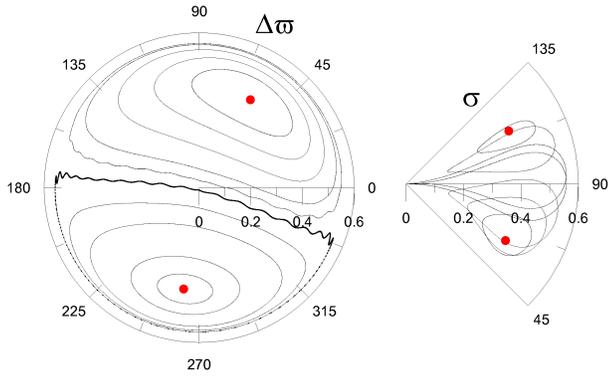}}
\caption{Variation of the eccentricity of each planet ($e_i$) with $\Delta\varpi$ (left frame) and with $\sigma$ (right frame) for initial conditions inside the stable region connecting $L_4$ and $AL_4$ (selected from line B in Fig. \ref{fig1} for $e=0.40$). The radial distance is the value of the osculating eccentricity. The resonant angle $\Delta\varpi$ oscillates around one of two possible centers. One corresponds to the $L_4$ configuration while the other to $AL_4$ configuration. Locations of the ACR solutions are given by the fixed points in red.}
\label{fig4}
\end{figure}

Table \ref{tab_pos} summarizes the detected stable stationary solutions in the planar planetary three-body problem, as well as their location in the plane of angular variables for low eccentricities.

\subsection{Motion Around the Stationary Solutions}

In order to visualize the dynamics of stable orbits outside the ACR, we integrated several orbits with initial elements $a_1=a_2=1$ AU, $e_1=e_2=0.4$, $\sigma=0$ and different values of $\Delta\varpi$. Each initial condition was chosen along line A drawn in Fig. \ref{fig1} for $e_i=0.40$. Results are shown in Figure \ref{fig3}. The left-hand plot shows the orbital evolution in the $(e_2,\Delta \varpi)$ plane, while the right-hand graph presents the variation of $(e_2,\sigma)$. In both cases the numerical output was filtered to eliminate short-period variations associated to the mean anomalies of both planets. Note that all trajectories display small-amplitude oscillations in $\sigma$, consistent with starting positions near the $\sigma$-family of periodic orbits. 

The behavior in the $(e_2,\Delta \varpi)$ plane is more intriguing. Initial conditions with $\Delta \varpi \in (90^\circ,270^\circ)$ exhibit oscillations of different amplitudes around the ACR solutions corresponding to quasi-satellite motion. Recall that this ACR solutions is located at $\Delta \varpi = 180^\circ$. However, initial conditions with $-90^\circ < \Delta\varpi < 90^\circ$  display regular motion that seem associated to large amplitude oscillations around $\Delta \varpi = 0$, even though this is an unstable point leading to close encounters and a collision between both planets. Nevertheless, there appears to be a minimum allowed amplitude for these solutions (shown in Figure \ref{fig3} as a blue dashed curve), which corresponds to a semi-amplitude in $\Delta \varpi$ of approximately $\sim 45^\circ$. Smaller amplitudes are unstable and lead to the ejection of one of the planets in short timescales.

Figure \ref{fig4} shows results for initial conditions inside the stable region connecting $L_4$ and $AL_4$. Semimajor axes and eccentricities were the same as in the previous plots. The initial values of $\Delta \varpi$ were varied from zero to $360^\circ$, and in each case $\sigma$ was chosen along line B in Fig. \ref{fig1} for $e_i=0.4$ \textbf{({\it $\sigma$-family})}. 

The $(e_2,\Delta \varpi)$ plane (left frame) shows two centers of oscillation, one corresponding to each ACR, and identified by red dots. $L_4$ is located at $\Delta \varpi=60^\circ$ while $AL_4$ roughly at $\Delta \varpi=240^\circ$. As before, we see a smooth transition in the dynamical behavior between both modes, with no evidence of any separatrix. Consequently, it appears that any initial condition will lead to a stable oscillation of $\Delta \varpi$ around the nearest stationary solution. 

The motion of the resonant angle $\sigma$ (right frame) shows a different behavior. Only initial conditions very close to either $L_4$ or $AL_4$ will show a small-amplitude circulation around the corresponding stationary point. As an example, notice some trajectories oscillating around $\sigma=90^\circ$ without reaching the fixed points. Finally, due to the intrinsic symmetry in co-orbital motion, the same behavior is also noted for initial conditions between $L_5$ and $AL_5$.

\begin{figure}
\centerline{\includegraphics*[width=18pc]{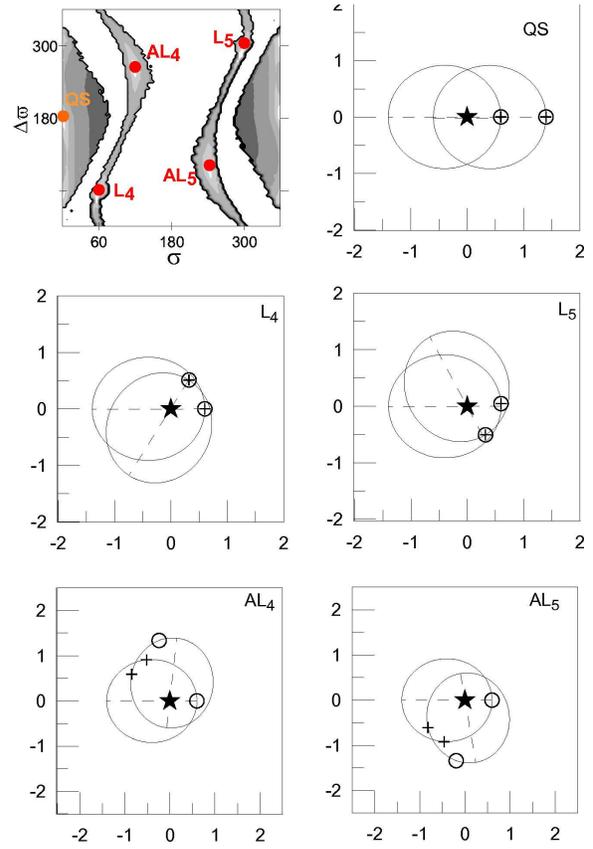}}
\caption{Orbit configuration with initial conditions chosen inside the regions of stable motions (see top left-hand frame). Initial position of planets are marked with open circles while crosses indicate the orbital configuration position leading to a minimum distance between the bodies. For QS, $L_4$ and $L_5$, the minimum distance coincides with the initial condition.}
\label{fig5}
\end{figure}

To better visualize each stable configuration, Figure \ref{fig5} presents the orbit scheme for five stable solutions, whose initial values of the angles are shown in the top left-hand frame. Five initial conditions correspond to the stable ACR solution discussed previously (QS, $L_4$, $L_5$, $AL_4$, $AL_5$). Each of the other plots show the orbital representation of each solutions in $(x,y)$ astrocentric cartesian coordinates. Initial conditions for both planets are shown in open circles, with $m_1$ located along the $x$-axis. Both axis directions are fixed. The orbital trajectory of each planet (over one period) is drawn in thin black lines, and the configuration leading to a maximum approach between both planets is shown with crosses. For QS, $L_4$ and $L_5$, the minimum distance coincides with the initial condition.

For QS orbits, the relative position of $m_2$ is always located in the positive $x$-axis, similar to the behavior noted in the restricted three-body problem (Mikkola et al. 2006). The relative motion of all five ACR solutions are periodic orbits, and symmetric with respect to the $x$-axis.

\section{Semi Analytical Model}\label{model}

One drawback in the previous numerical approach is the excessive CPU time required for the construction of each dynamical map. In order to extended these results to other values of the parameter space (e.g. planetary masses, eccentricities) it is useful to construct a semi-analytical model for the co-orbital motion. 

Such a model can be developed along similar lines as other mean-motion resonances 
(e.g. Michtchenko et al, 2006, 2008a, b). It requires two main steps: first, a transformation to adequate resonant variables and, second, a numerical averaging of the Hamiltonian with respect to short-period terms. Both tasks are detailed below.

We begin introducing the usual mass-weighted Poincar\'e canonical variables (e.g Laskar 1990) for each planet $m_i$:
\bea
\begin{array}{rcll}
\lambda_i & ; & L_i = m'_i \sqrt{\mu_i a_i} & × \\ 
\varpi_i  & ; & G_i-L_i = -L_i\left(1-\sqrt{1-e_i^2}\right) & ×
\end{array}
\eea
where $\mu_i = \kappa^2(m_0+m_i)$, $\kappa$ denotes the gravitational constant, and $m_i'$ is the reduced mass of each body, given by:
\bea
m_i' = {m_i m_0 \over m_i+m_0}.
\eea
The Hamiltonian function $F$ can be expressed as $F = F_0 + F_1$, where $F_0$ corresponds to the two-body contribution, and has the form:
\bea
F_0 = -\sum_{i=1}^2{\mu_i^2m_i'^3 \over 2L_i^2}.
\eea
The second term, $F_1$, is the disturbing function which can be written as:
\bea
F_1=-\kappa^2m_1m_2 {1\over \Delta} + T_1, 
\eea
where $\Delta$ is the instantaneous distance between the two planets, and $T_1$ is the indirect part of the potential energy of the gravitational interaction (see Laskar 1990, Laskar and Robutel 1995 for more details).

\begin{figure}
\centerline{\includegraphics*[width=20pc]{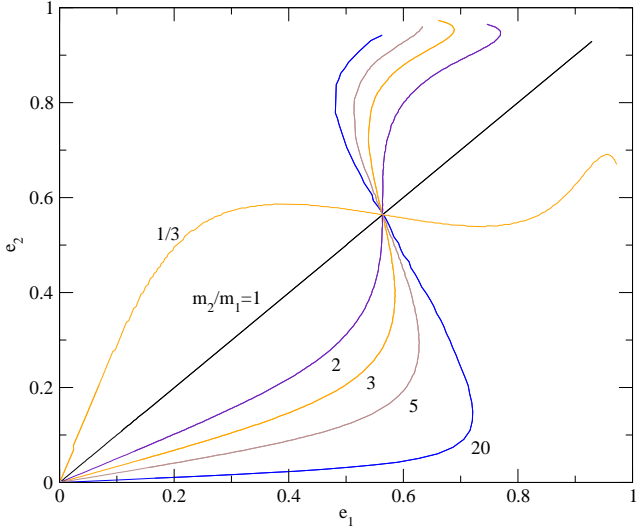}}
\caption{Families of stable QS stationary solutions in the $(e_1,e_2)$ plane, for  three different mass ratios $m_2/m_1$. Notice a locus of solutions at $e_1=e_2=0.565$ for all masses. The solutions for $m_2/m_1<1$ are mirror images of those for $m_2/m_1>1$.}
\label{fig6} 
\end{figure}

For initial conditions in the vicinity of co-orbital motion, we define the following set of planar resonant canonical variables $(I_1,I_2,{\cal K},{\cal AM},\sigma,\Delta \varpi,Q,q)$, where:
\bea
\label{canour}
\begin{array}{lcl}
\;\;\;\; \sigma=\lambda_2-\lambda_1   & ; & \;\;\;\; I_1=\frac{1}{2}(L_2-L_1) \\ 
\Delta\varpi=\varpi_2-\varpi_1        & ; & \;\;\;\; I_2=\frac{1}{2}(G_2-G_1-L_2+L_1) \\ 
\;\;\;\; q=\varpi_2+\varpi_1          & ; & \;\;\;\; J_1=\frac{1}{2}(G_1+G_2) \\ 
\;\;\; Q=\lambda_1+\lambda_2-q        & ; & \;\;\;\; J_2=\frac{1}{2}(L_1+L_2) \end{array}
\eea
where, $J_1=\frac{1}{2}{\cal AM}$ and $J_2=\frac{1}{2}{\cal K}$. A generic argument $\varphi$ of the disturbing function can be written as:
\be
\label{pphi}
\varphi = j_1\lambda_1+j_2\lambda_2+j_3\varpi_1+j_4\varpi_2,
\ee
where $j_k$ are integers. In terms of the new angles the same argument may be written as:
\be
\varphi = \frac{1}{2}\left[ (j_2-j_1)\sigma + (j_4-j_3)\Delta\varpi + (j_1+j_2)Q \right].
\ee
Since $q$ is a cyclic angle, the associated action ${\cal AM}$ is a constant of motion (total angular momentum) of the system.

The next step is an averaging of the Hamiltonian over the fast angle $Q$. This procedure can be performed numerically, allowing to evaluate the averaged Hamiltonian $\bar{F}$ as:
\be
\label{aver}
\bar{F}(I_1,I_2,\sigma,\Delta\varpi;{\cal K},{\cal AM}) \equiv \frac{1}{2\pi} \int_0^{2\pi} F dQ. 
\ee
In the averaged variables, ${\cal K}$ is a new integral of motion which, in analogy to other mean-motion resonances (e.g. Michtchenko et al. 2008a), we call the {\it scaling parameter}. 

$\bar{F}$ then constitutes a system with two degrees of freedom in the canonical variables ($I_1,I_2,\sigma, \Delta\varpi$), parametrized by the values of both ${\cal K}$ and ${\cal AM}$. Since the numerical integration depicted in equation (\ref{aver}) is equivalent to a first-order averaging of the Hamiltonian function (e.g. Ferraz-Mello, S. 2007), only those periodic terms (\ref{pphi}) with $j_1+j_2=0$ remain in $\bar{F}$. In consequence, we can rewrite the generic resonant argument of the averaged system as:
\be
\varphi = j_2\sigma + j_4\Delta\varpi.
\label{eqperiodic0}
\ee
where the index $j_2,j_4$ are integers that may take any value in the interval $(-\infty,\infty)$.

\section{Families of Periodic Orbits}\label{op}

In the averaged system defined by $\bar{F}$ exact zero-amplitude ACR solutions are given by the stationary conditions:
\bea
\label{eq_hamil}
\frac{\partial \bar{F}}{\partial \sigma}= \frac{\partial \bar{F}}{\partial \Delta\varpi}= \frac{\partial \bar{F}}{\partial I_1}= \frac{\partial \bar{F}}{\partial I_2}=0 ,
\eea
and can therefore be identified as extrema of the averaged Hamiltonian function. In the present section we will use this approach to estimate the families of different ACR as function of the planetary masses and eccentricities, and compare the results with numerical integrations of the exact equations of motion.

\begin{figure}
\centerline{\includegraphics*[width=20pc]{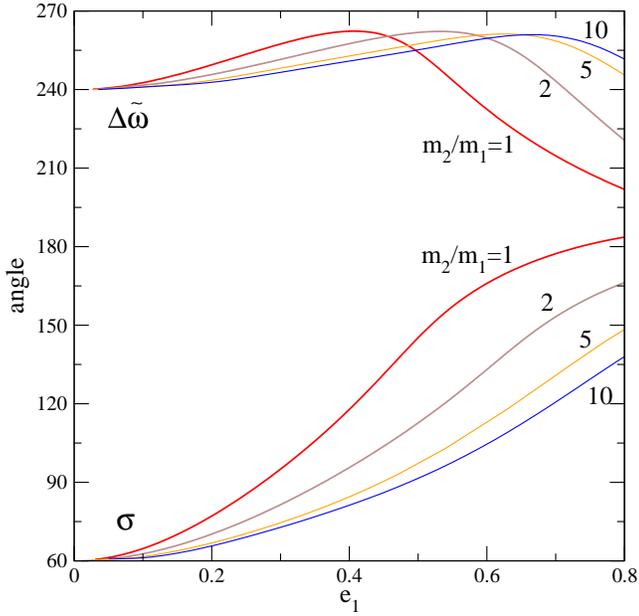}}
\caption{Equilibrium values of $\sigma$ and $\Delta \varpi$ for the family of $AL_4$ solutions as function of the eccentricity of the smaller planet, for several mass ratios $m_2/m_1 \ge 1$.}
\label{fig7}
\end{figure}

\subsection{Families of Symmetric ACR. QS}

We begin calculating the exact stationary solutions, corresponding to QS configurations, as a function of the eccentricities, and for different values of the planetary masses. As mentioned in Hadjidemetriou et al. (2009), the locations and stability of the ACR do not appear dependent on the individual values of the masses, but only on their ratio $m_2/m_1$.

\begin{figure}
\centerline{\includegraphics*[width=20pc]{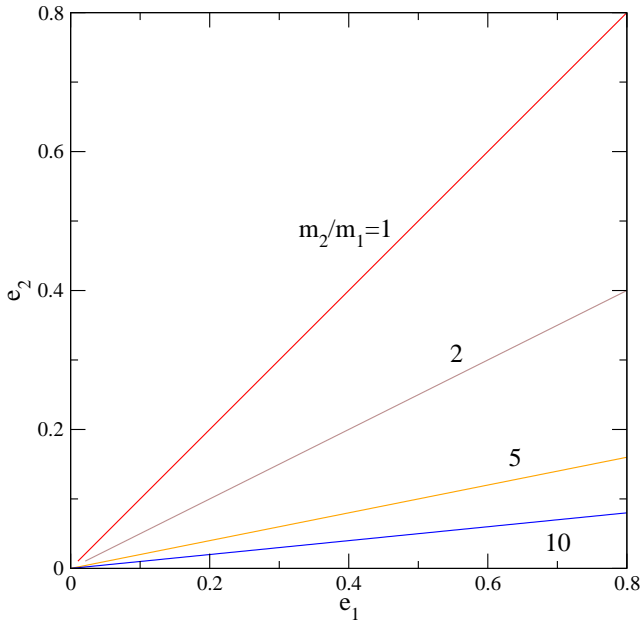}}
\caption{Families of $AL_4$ solutions, $e_2$ as function of $e_1$, for several mass ratios $m_2/m_1 \ge 1$.}
\label{fig8} 
\end{figure}

In all cases, the stationary values of the canonical momenta $L_i$ are such that
$n_1=n_2$, where $n_i$ are the mean motions of the planets. For equal mass planets, this reduces to the condition $a_1=a_2$. Finally, the angles of the exact ACR always remain locked at $(\sigma,\Delta\varpi)=(0,180^\circ)$. Hadjidemetriou et al (2009) presented similar plots for the same mass ratios. 

Figure \ref{fig6} shows the families of stable zero-amplitude QS orbits for selected mass ratios: $m_2/m_1 = 1/3$, $m_2/m_1 = 1$, and $m_2/m_1 > 1$. For equal masses, all solutions occur for and $e_1=e_2$. Due to the intrinsic symmetry of the dynamical system, the family of stationary solutions for $m_2/m_1 = 1/3$ is a mirror image of the solution for $m_2/m_1 = 3$, since it may be obtained by simply interchanging $e_1$ with $e_2$. In the case of $m_2/m_1 = 3$, we note that $e_2 < e_1$ for $e_2 < 0.565$, while $e_2 > e_1$ for more elliptic orbits.
Figure \ref{fig6} shows also the solutions for $m_2/m_1 = 2,\;5,\;20$ mass ratios. For mass ratios smaller than unity, the solutions are mirror images with respect to the family $m_2/m_1 = 1$. Note that the families of stable solutions approach $e_1=e_2$ as $m_2 \rightarrow m_1$. However, as the mass ratio tends towards the restricted three-body problem, the eccentricity of the smaller mass approaches unity. Finally, the solution $e_i=0.565$ is common to all the QS families, and corresponds to a global extrema of the Hamiltonian in this plane. A similar structure was already noted by Michtchenko et al. (2006) for other mean-motion resonances.

\subsection{Families of Asymmetric ACR Solutions. $L_4$ and $AL_4$}

The same procedure can also be applied to the Lagrangian $L_4$ and Anti-Lagrangian $AL_4$ configurations. Recall that the dynamical maps (Figure \ref{fig2}) 
showed a symmetry with respect to the transformation $(\sigma,\Delta \varpi) \rightarrow (-\sigma,-\Delta \varpi)$, so the results discussed here can also be applied to the $L_5$ and $AL_5$ solution, by applying the same operation on the variables.

The ACR solution associated to the Lagrangian solution $L_4$ shows no variation in the angles, maintaining constant both angles at $60^\circ$. The solutions remain stable for initial conditions up to eccentricities $e_i=0.7$. 
 However, the $AL_4$ shows significant changes as function of the eccentricities. Figure \ref{fig7} shows the equilibrium values of both angles for the family of $AL_4$, as a function of the eccentricity of the smallest planet, for several values of the mass ratio $m_2/m_1$.  The resonant angle $\sigma$ increases monotonically from $60^\circ$, at quasi-circular orbits, towards $\sim 180^\circ$ for near parabolic trajectories. As the mass ratios increases, the maximum value of the resonant angle decreases, reaching $\sigma=150^\circ$ for a mass ratio of $m_2/m_1=10$. 

The secular angle $\Delta \varpi$ shows a slightly more complex behavior. Initially it increases from $ \sim 240^\circ$ until it reaches a maximum value close to $\sim 260^\circ$, after which it once again decreases towards  $\Delta \varpi \sim 180^\circ$. The planetary eccentricity corresponding to the maximum in the secular angle increases with the mass ratio, approaching the parabolic limit for $m_2/m_1 \sim 10$. 

As shown in Figures \ref{fig1} and \ref{fig2}, the size of the stable region around each asymmetric solution decreases with the increase of $e_i$, and practically disappears as the angles approach $180^\circ$ degrees. For quasi-parabolic orbits, only the region around $AL_4$ is discernible. Thus, for high eccentricity planets in co-orbital motion, it appears that the $AL_4$ and $AL_5$ asymmetric solutions are more regular than the classical equilibrium Lagrangian solutions $L_4$ and $L_5$. 

The values of the planetary eccentricities at $AL_4$ for different mass ratios is presented in Figure \ref{fig8}. Contrary to the QS trajectories, there appears to be a purely linear dependence between $e_2$ and $e_1$ as a function of the mass ratio. In fact, a simple numerical analysis of the results appears to indicate that 
\be
e_1 \simeq \biggl( \frac{m_2}{m_1} \biggr) e_2.
\label{e1e2}
\ee
Thus, for mass ratios approaching the restricted three-body problem (with $m_2 \rightarrow 0$) it should be expected that the eccentricity of the massive planet $m_1$ at the $AL_4$ solution would tend towards zero. 

\begin{figure}
\centerline{\includegraphics*[width=20pc]{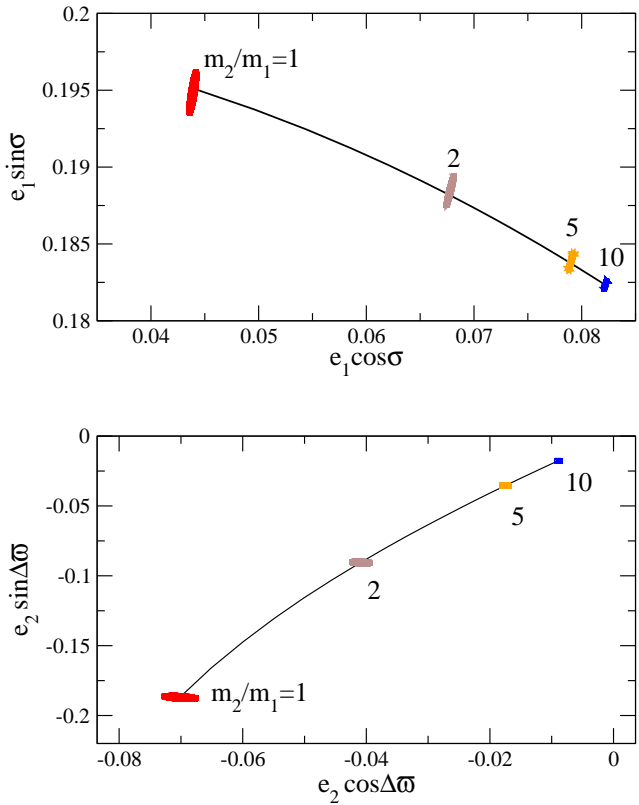}}
\caption{Exact numerical integrations of initial conditions close to the $AL_4$  stationary solutions ($m_2/m_1 \ge 1$). Black curves represent the $AL_4$-family of ACR calculated with the semi-analytical model.}
\label{fig9} 
\end{figure}

Finally, the equilibrium values of the semimajor axes also change as function of the mass ratio. Here, however, it is easy to see from the stationary conditions 
(\ref{eq_hamil}) that a zero-amplitude $AL_4$ trajectory is characterized by the relation $n_1 = n_2$. For equal mass planets, this reduces to $a_1=a_2$. 

The families of stationary solutions presented in this section were calculated using our semi-analytical model. In order to compare them with actual numerical simulations of the exact equations, we choose four solutions from Figure \ref{fig7} with $e_1=0.2$, but corresponding to different mass ratios. Each was then numerically integrated for several orbital periods, assuming zero initial values for the cyclic angular variables $q$ and $Q$. Results are shown in Figure \ref{fig9}, where the top frame presents the trajectories in the plane $(e_1 \cos{\sigma}, e_1 \sin{\sigma})$ and the bottom graph in the plane $(e_2 \cos{\Delta \varpi}, e_2 \sin{\Delta \varpi})$. Each initial condition shows a small amplitude oscillation around the stationary value, which presents a very good agreement with the family of $AL_4$ solutions calculated with our model (black curve). 
\begin{figure}
\centerline{\includegraphics*[width=20pc]{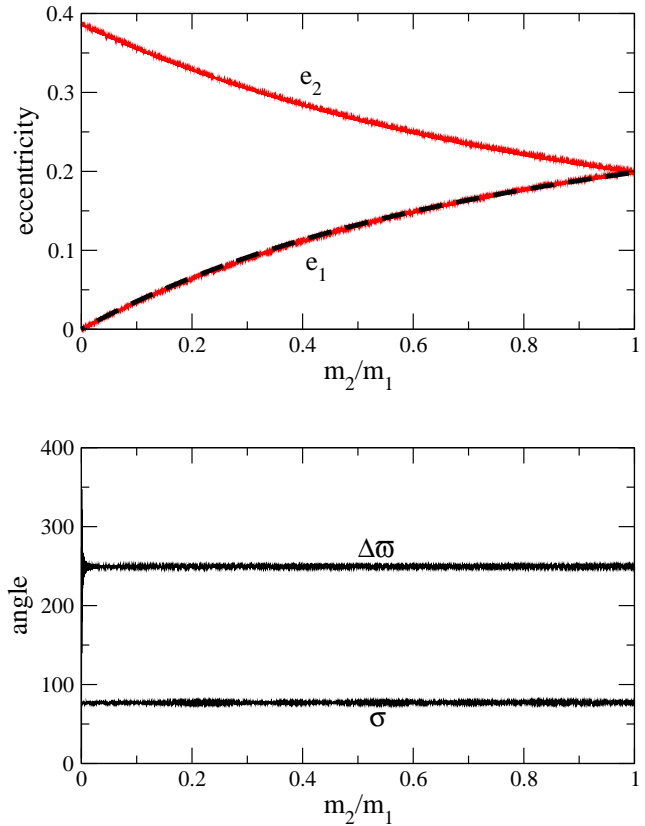}}
\caption{Evolution of $AL_4$ applying a smooth adiabatic decrease in $m_1$. Initial conditions correspond to $m_2=m_1$ and $e_1=e_2=0.2$. The stationary values of the angles are invariant to the mass change, although the amplitude of $\Delta \varpi$ increases as $e_2$ approaches zero. The ratio of the equilibrium eccentricities follow the relation (\ref{e1e2}), as shown by the dashed black curve overlaying the data of $e_2$.}
\label{fig10} 
\end{figure}

\subsection{Adiabatic Mass Variation in $AL_4$}

As a final analysis, in this section we study the orbital evolution of a system initially near $AL_4$, when the mass of one of the planets is decreased adiabatically. This question is raised for three reasons. First, as shown by Lee (2004), for two planets in a 2/1 mean-motion resonance, a sufficiently slow change in one of the masses will preserve the resonant configuration and allow to calculate the variation of the ACR as a function of $m_2/m_1$. In other words, this approach provides a different numerical test of our semi-analytical model and an alternative way to calculate the stationary orbits. Second, the results will also allow us to test the robustness of the new asymmetric co-orbital solutions $AL_4$ and see how they respond to changes in the parameters of the system. Finally, we wish to analyze the behavior of these new solutions in the limit of the restricted three-body problem, corresponding to $m_2=0$.

Figure \ref{fig10} shows a typical example. Initial conditions correspond to an $AL_4$ solution for $m_2/m_1 = 1$ and $e_i = 0.2$. While $m_1$ was maintained fixed, $m_2$ was varied linearly down to $m_2=0$ in a timescale of $10^6$ orbital periods. We checked using other timescales, finding no significant variations. This guarantees that we are effectively in the adiabatic regime. 

The top graph of Figure \ref{fig10} shows the evolution of the orbital eccentricities as function of the mass ratio. As soon as $m_2/m_1$ departs from unity, the value of $e_2$ increases while $e_1$ decreases.
The broken black curve that can be seen over the red curve shows the predicted value of $e_1$ applying the relation (\ref{e1e2}) to each value of $e_2$. The agreement is excellent, giving an additional corroboration to this empirical relationship between the eccentricities. It must be noted that neither the total angular momentum ${\cal AM}$ nor the scaling parameter ${\cal K}$ are preserved during the mass change.
The bottom plot of Figure \ref{fig10} shows the behavior of the angular values during the mass variation. The equilibrium values of both $\sigma$ and $\Delta \varpi$ remain practically unchanged. 

\begin{figure}
\centerline{\includegraphics*[width=20pc]{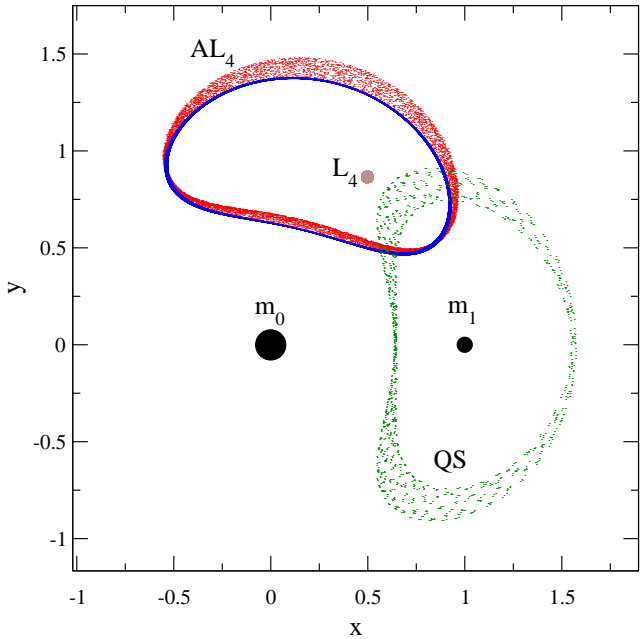}}
\caption{Orbital trajectories in a cartesian rotating pulsating reference frame, where the positions of $m_0$ and $m_1$ are fixed in the $x$-axis with unit mutual distance. Both bodies are marked with large filled circles. The brown circle shows the orbital evolution of $m_2$ when placed in $L_4$, while red dots correspond to an initial condition for $m_2$ placed in $AL_4$ (see Figure \ref{fig10}). In both cases the plot presents the orbital evolution as $m_2\rightarrow 0$. No change is observed in $L_4$, while the blue curve shows the final orbital trajectory around $AL_4$ when $m_2$ reaches zero. The oscillation period is equal to the orbital period between the massive primaries.}
\label{fig11} 
\end{figure}

For smaller mass ratios $m_1$ tends towards a circular orbit, while the eccentricity of the smaller planet approaches $e_2 \sim 0.4$. This seems to imply that the asymmetric $AL_4$ (and consequently $AL_5$) solutions could also exist in the limit of the restricted three-body problem. To test this conjecture and compare the trajectories of both $L_4$ and $AL_4$ solutions in the restricted ($m_2 \rightarrow 0$) limit, Figure \ref{fig11} plots the $(x,y)$ cartesian coordinates of two initial conditions in a rotating pulsating reference frame. 

In the rotating pulsating reference frame the positions of both $m_0$ and $m_1$ are fixed in the $x$-axis (shown with large solid circles). Three orbital evolutions are shown: the brown dot corresponds to initial conditions in the asymmetric $L_4$, while red dots map the evolution of an orbit originally in $AL_4$. In both cases we started with $m_2=m_1$, but subsequently decreased $m_2$ to zero (restricted case). No change is noticed in the $L_4$ orbit, and the trajectory remained in an equilateral configuration with the two finite masses. However, the $AL_4$ solution converged towards a tadpole-type orbit of large amplitude (blue curve) for $m_2 \rightarrow 0$. This solution corresponds to a periodic orbit whose period coincides with the orbital period of the primaries around the center of mass. Green dots map the evolution of an orbit originally in QS. As we can see the orbit described by QS configuration revolves around the $m_1$, in the same way that was observed in the restricted problem.

Thus, there appears to be a structural difference between the $L_4$ and $AL_4$ planetary solutions discussed in this paper. Although both appear as ACR (fixed points in the averaged problem) the first are true stationary solutions in the unaveraged rotating frame, while the new solutions $AL_4$ are actually large amplitude periodic orbits that encompass the classical Lagrangian equilateral solution. 

\section{Conclusions}\label{conclusions}

We studied the stability regions and families of periodic orbits of two-planet systems in the vicinity of a 1/1 mean-motion resonance (i.e. co-orbital configuration). We considered different ratios of planetary masses and orbital eccentricities, also we assumed that both planets share the same orbital plane (coplanar motion). 

As result we identified two separate regions of stability, each with two distinct modes of motion:

\begin{itemize}
\item 
\textbf{Quasi-Satellite region:} Originally identified by Hadjidemetriou et al. (2009) for the planetary problem, QS orbits correspond to oscillations around an ACR located at $(\sigma,\Delta\varpi) = (0,180^{\circ})$. Although not present for quasi-circular trajectories, they fill a considerable portion of the phase space in the case of moderate to high eccentricities.

We also found a new regime, associated to stable orbits displaying oscillations around $(\sigma,\Delta\varpi) = (0,0)$, even though this point is unstable and corresponds to a collision between the two planets. 

\item 
\textbf{Lagrangian region:} Apart from the previous symmetric solutions, we also found two distinct types of asymmetric ACR orbits in which both $\sigma$ and $\Delta\varpi$ oscillate around values different from $0$ or $180^\circ$. The first is the classical equilateral Lagrangian solution associated to local maxima of the averaged Hamiltonian function. Independently of the mass ratio $m_2/m_1$ and their eccentricities, these solutions are always located at $(\sigma,\Delta\varpi) = (\pm 60^\circ,\pm 60^\circ)$. However, the size of the stable domain decreases rapidly for increasing eccentricities, being practically undetectable for $e_i > 0.7$.

The second type of asymmetric ACR correspond to local minima of the averaged Hamiltonian function. We have dubbed them Anti-Lagrangian solutions ($AL_4$ and $AL_5$). For low eccentricities, they are located at $(\sigma,\Delta\varpi) = (\pm 60^\circ,\mp 120^\circ)$. Each is connected to the classical $L_4$ and $L_5$ solution through the $\sigma$-family of periodic orbits in the averaged system. Contrary to the classical equilateral Lagrangian solution, their location in the plane $(\sigma,\Delta\varpi)$ varies with the planetary mass ratio and eccentricities. Although their stability domain also shrinks for increasing values of $e_i$ they do so at a slower rate than the classical Lagrangian solutions, and are still appreciable for eccentricities as high as $\sim 0.7$. 
\end{itemize}

Finally, we also applied an ad-hoc adiabatically slow mass variation to one of the planetary bodies, and analyzed its effect on the $AL_4$ configuration. We found that the resonant co-orbital solution was preserved, with practically no change in the equilibrium values of the angles. The eccentricities, however, varied with the larger planet approaching a quasi-circular orbit as the smaller planet had its eccentricity increased. These solution still exist in the limit of the restricted three-body problem (i.e. $m_2 \rightarrow 0$), although both types of asymmetric solutions ($L_4$ and $AL_4$) have different geometries. While the first are true stationary solutions in the unaveraged system, the latter are periodic orbits around the classical equilateral Lagrangian points.

\section*{Acknowledgments}
This work has been supported by the Argentinian Research Council -CONICET-, the Brazilian National Research Council -CNPq-, and the S\~ao Paulo State Science Foundation -FAPESP-. The authors also gratefully acknowledge the CAPES/Secyt program for scientific collaboration between Argentina and Brazil.

\label{lastpage}


\begin{thebibliography}{}

\bibitem[]{bfm03}
Beaug\'e C., Ferraz-Mello S., Michtchenko T. A., 2003, ApJ, 593, 1124.

\bibitem[]{Brass04}	
Brasser, R., Innanen, K. A., Connors, M., Veillet, C., Wiegert, P., Mikkola, S., Chodas, P.W., 2004, Icarus, 171, 102.

\bibitem[]{cin00}
Cincotta P.M., Sim\'o C., 2000, A\&AS, 147, 205.

\bibitem[]{con02}
Connors M., Chodas P., Mikkola S.,Wiegert P., Veillet C., Innanen K., 2002,
Meteoritics Planet. Sci., 37, 1435.

\bibitem[]{fm07} Ferraz-Mello, Sylvio, 2007, Canonical Perturbation Theories: Degenerate Systems and Resonance. Astrophysics and Space Science Library, Vol. 345. Springer, NY.

\bibitem[]{gk06}
Go\'zdziewski K., Konacki M., 2006, ApJ, 647, 573.

\bibitem[]{had09}
Hadjidemetriou J., Psychoyos D., Voyatzis G., 2009, CeMDA, 104, 23.

\bibitem[]{h69}
H\'enon, M., 1969, A\&A, 1, 223.

\bibitem[]{j13}
Jackson, J., 1913, MNRAS, 74, 62.


\bibitem[]{kor05}
Kortenkamp S., 2005, Icarus, 175, 409.

\bibitem[]{l90}
Laskar, J., 1990, In D. Benest, C. Froeschl\'e (eds.) {\it Les M\'ethodes Modernes de la M''ecanique C\'eleste} (Goutelas 89). 

\bibitem[]{lr95}
Laskar, J., Robutel, Ph., 1995, CeMDA, 62, 193.

\bibitem[]{lee04}
Lee, M.H., 2004, ApJ, 61, 784.


\bibitem[]{mbf06}
Michtchenko, T.A., Beaug\'e, C., Ferraz-Mello, S., 2006, CeMDA, 94, 411.

\bibitem[]{mbf08a}
Michtchenko, T.A., Beaug\'e, C \& Ferraz-Mello, S., 2008a. MNRAS, 387, 747.

\bibitem[]{mbf08b}
Michtchenko, T.A., Beaug\'e, C \& Ferraz-Mello, S., 2008b. MNRAS, 391, 227.

\bibitem[]{mik97}
Mikkola, S., Innanen, K., 1997, In The Dynamical Behavior of our Planetary System, ed., R. Dvorak and J. Henrard (Dordrecht: Kluwer), 345.

\bibitem[]{mikk04}
Mikkola, S., Brasser, R., Wiegert, P., Innanen, K., 2004, MNRAS, 351, 63

\bibitem[]{mik06}
Mikkola S., Innanen K., Wiegert P., Connors M., Brasser R., 2006, MNRAS, 369, 15.

\bibitem[]{nam09}
Namouni, F., 1999, Icarus, 137, 293.


\bibitem[]{wim00}
Wiegert, P., Innanen K., Mikkola S., 2000, AJ, 119, 1978.

\end{thebibliography}
\end{document}